\documentclass[runningheads]{llncs}

\usepackage{bibnames}
\usepackage{soul,xcolor}
\usepackage{subfloat}
\usepackage{subcaption}
\usepackage{graphicx} 
\usepackage{newtxtext,newtxmath}
\usepackage{url}
\usepackage[mathscr]{eucal}
\usepackage{amsmath,amsfonts}
\usepackage{siunitx}        
\usepackage{etoolbox}   
\usepackage{booktabs}

\usepackage{amsmath}
\usepackage{cite}
\usepackage{multirow}

\newcommand{\ie}{\emph{i.e.~}}

\begin{document}

\title{Learning to Recommend Items to Wikidata Editors}

\author{Kholoud AlGhamdi \and
Miaojing Shi \and
Elena Simperl}

\authorrunning{K. AlGhamdi et al.}

\institute{King's College London, United Kingdom\\ 
\email{\{kholoud.alghamdi, miaojing.shi, elena.simperl\}@kcl.ac.uk}\\ }
\maketitle              

\begin{abstract}

Wikidata is an open knowledge graph built by a global community of volunteers. As it advances in scale, it faces substantial challenges around editor engagement. These challenges are in terms of both attracting new editors to keep up with the sheer amount of work, and retaining existing editors. Experience from other online communities and peer-production systems, including Wikipedia, suggests that personalised recommendations could help, especially newcomers, who are sometimes unsure about how to contribute best to an ongoing effort. For this reason, we propose a recommender system \emph{WikidataRec} for Wikidata items. The system uses a hybrid of content-based and collaborative filtering techniques to rank items for editors relying on both item features and item-editor previous interaction. A neural network, named neural mixture of representations, is designed to learn fine weights for the combination of item-based representations and optimize them with editor-based representation by item-editor interaction.  
To facilitate further research in this space, we also create two benchmark datasets, a general-purpose one with $220,000$ editors responsible for $14$ million interactions with $4$ million items, and a second one focusing on the contributions of more than $8,000$ more active editors. We perform an offline evaluation of the system on both datasets with promising results. Our code and datasets are available at https://github.com/WikidataRec-developer/Wikidata\_Recommender.


\end{abstract}
%
%
%
\section{Introduction}
\label{sec:introduction}
Wikidata is an open knowledge graph built by a global community of volunteers \cite{vrandevcic2014wikidata}. Since its launch in $2012$ it reached more than $90$ million items and $24,000$ active editors\footnote{\url{www.wikidata.org/wiki/Wikidata:Statistics/en}}. Manual contributions are core to Wikidata \cite{vrandevcic2014wikidata}: editors add new content, keep it up-to-date, model knowledge as graph items and properties, and decide on all rules of content creation and management. However, as Wikidata advances in scale, it faces substantial challenges around editor engagement \cite{sarasua2019evolution}. 

Experience from other online communities and peer-production systems, including Wikipedia, Quora and others, suggests that one way to improve engagement is through \textit{recommendations} \cite{dror2011want,freyne2009increasing,yang2016recommending}. This is especially true for editors who are relatively new to Wikidata, who need to overcome so-called `legitimate peripheral participation' (LPP) effects to continue to engage \cite{piscopo2017wikidatians}. According to LPP, newcomers are more likely to become members of a community if they are provided with suggestions of (typically lower-risk) tasks they could carry out to further the goals of the community \cite{lave1999legitimate}.

At the moment, looking for relevant items to edit in Wikidata remains challenging because of the sheer number of options available \cite{sarasua2019evolution}. The suggested and open tasks page \footnote{\url{www.wikidata.org/wiki/Wikidata:Contribute/Suggested_and_open_tasks}} gives a useful but daunting overview of the various ways in which people could contribute to Wikidata. Many task lists are automatically generated, without taking into account aspects such as editor tenure, previous editing history or interests. This lack of focus is also said to prevent editors from developing reinforced editing habits, which increases the likelihood of dropouts \cite{sarasua2019evolution}. Seasoned editors use tools such as \textit{QuickStatements}\footnote{\url{https://quickstatements.toolforge.org/#/}} and \textit{Watchlist}\footnote{\url{www.wikidata.org/wiki/Help:Watchlist}} to organise their work, but such tools become relevant only once editors have identified productive ways to contribute.

For these reasons, we propose \textit{WikidataRec}, a recommender system for Wikidata items. The system uses content-based and collaborative filtering techniques to rank items for editors relying on both item features and item-editor previous interaction. Collaborative filtering is a representative approach to address recommendation task with implicit feedback, which learns item-centric and editor-centric edit representations using matrix factorization~\cite{koren2009matrix} from item-editor interaction data. We adopt it in this work and further facilitate it with additional information from item content and relations, where we learn these representations from by sentence embedding model ELMo~\cite{peters2018deep} and graph embedding model TransR~\cite{lin2015learning}, respectively. To combine the multiple representations, we introduce the neural mixture of representations (NMoR), a neural network inspired by mixture of experts~\cite{jacobs1991adaptive} that produces fine weights for different item-based representations. It is optimized over the editor-based representation to rank items to editors.  The proposed \textit{WikidataRec} demonstrates a large capacity of leveraging items' contents, relations, and edit history between editors and items into the recommender. 

To facilitate further research in this space, we also create two benchmark datasets, a general-purpose one with $220,000$ editors responsible for $14$ million interactions with $4$ million items, and a more focused one with contributions of more than $8,000$ active editors with more than $200$ edits each. We evaluate the system on these datasets against several baselines and analyse the impact of different features and levels of data sparsity on performance. The results are promising, though challenges remain because of unbalanced participation, sparse interaction data and little explicit information about  editors' interests and feedback.

\section{Background and Related Work}
\label{sec:background}

\subsection{Wikidata Data Model}
\label{sec:data_model}
\begin{figure}[t]
    \includegraphics[width=0.7 \columnwidth]{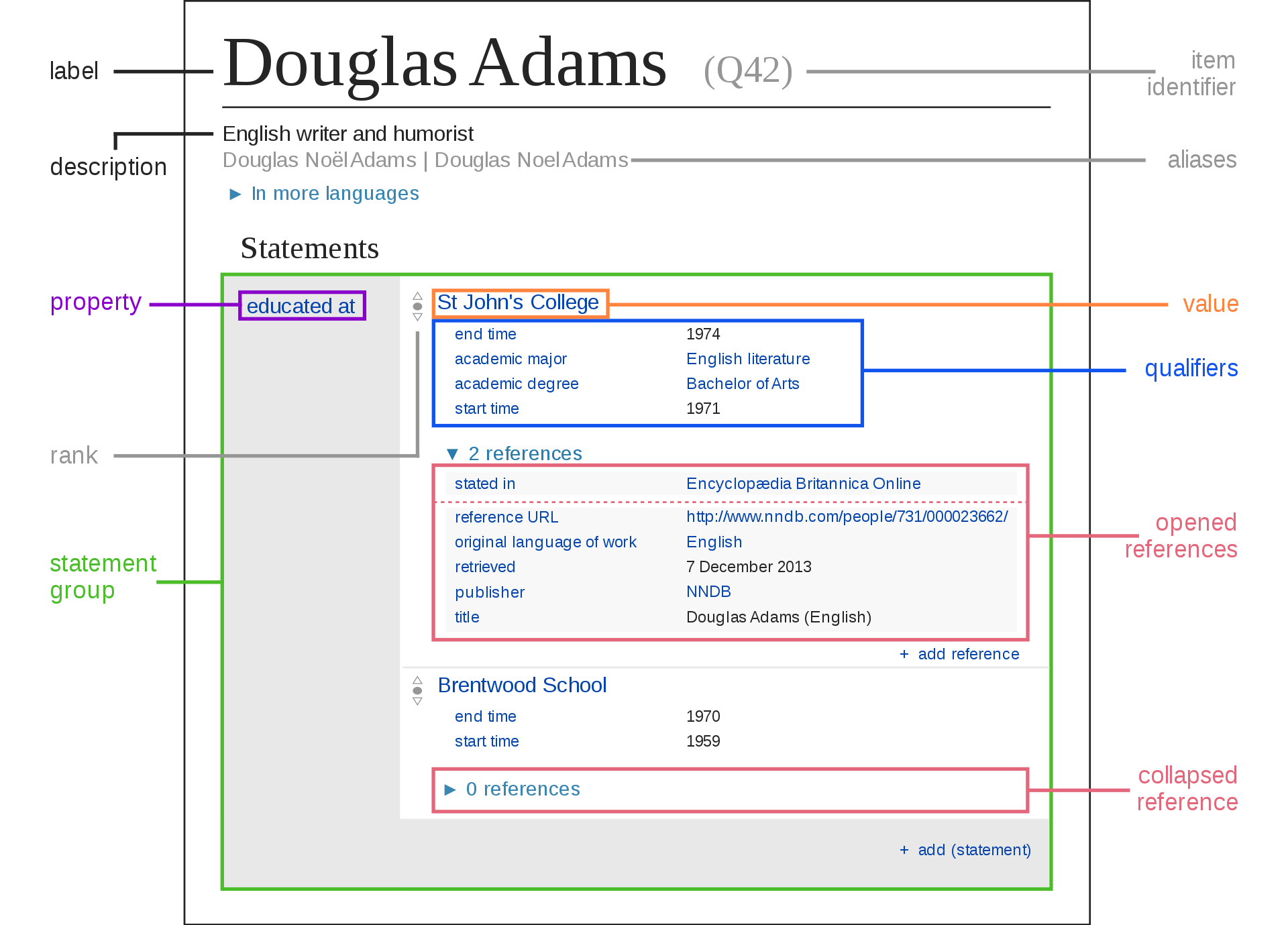}
    \centering 
    \caption{The structure of an Wikidata item. \\ Source: \cite{bworld}}
    \label{fig:wikidataexample}
    \vspace{-4mm}
\end{figure}

Wikidata consists of structured records stored in form of entities, where each entity is allocated a separate page and described by a set of terms and statements \cite{vrandevcic2014wikidata}. There are two types of entities, items and properties, situated in different namespaces. The namespaces are identified by URIs using numbers and letters, with the letter $Q$ relating to items and the letter $P$ to properties. Although there are other namespaces in Wikidata, we focus only on these two, as they are most relevant for our recommender system. Items and property pages have a similar interface, starting with labels, descriptions and aliases, which are language-specific, and then moving on to statements, which are language-agnostic. Each statement contains a set of triplet edges (head, relation, tail), capturing the different types of relations between entities in the world, e.g., London $\overrightarrow{\text{\scriptsize{capital of}}}$ United Kingdom. Pages also include sitelinks connecting to Wikipedia articles or other Wikimedia projects \cite{muller2015peer}. Each page is indexed by a unique identifier. 
Figure \ref{fig:wikidataexample} shows an example of a Wikidata item page. 

\vspace{-3mm}
\subsection{Editors and Editing in Wikidata} 
\label{sec:Wikidata-editing}

Anyone can edit Wikidata. The literature tends to distinguish between two types of contributions: manual ones, carried out by registered or anonymous editors, and bots. Bots are used for repetitive editing tasks, but do not contribute to discussions, modeling decisions, or rules for content creation, and management \cite{muller2015peer}. Human editors may get involved in any type of activities, though some require specific access rights or ontology engineering skills \cite{piscopo2017wikidatians,piscopo2018models}. 
Furthermore, there are two types of editing tasks which are higher-risk tasks and lower-risk ones. Edits are higher-risk if they affect a larger share of the graph, e.g. property editing; hence not all editors may create properties in Wikidata. Lower-risk edits are, for example adding/changing labels, descriptions, etc. Formally, the Wikidata editing process starts when an editor logs into the platform and contributes by inserting data on an item page using the basic editing interface (See Figure \ref{fig:wikidataexample}). Every action performed by an editor is recorded on the so-called revision history page. 
In our current work, we focus on item edits rather than properties or other types of contributions. Our aim is to establish whether recommendations are technically feasible with the available data. Previous studies into the Wikidata community suggest that item edits are popular with less experienced editors, who are the main audience for personalised recommendations. Properties or other sorts of works are normally dealt by seasoned and active editors \cite{piscopo2017wikidatians,piscopo2018models}.

\vspace{-3mm}
\subsection{Recommending Tasks to Communities and Crowds Online}
\label{sec:related-works}
There are a lot of works proposing the use of personalized recommendations in online communities and peer-production systems \cite{dror2011want,freyne2009increasing,yang2016recommending}. Recommendations aim to make work more effective and increase retention, by matching open tasks to people's skills and interests \cite{cosley2007suggestbot}. As noted in the introduction, they can also help new members of a community find their way and contribute~\cite{freyne2009increasing,piscopo2017wikidatians}.

In Wikipedia, a first recommender system, SuggestBot, was proposed in \cite{cosley2007suggestbot}. It used article titles, links, and co-editing patterns as main features to recommend articles to editors. A more recent recommender system by \cite{moskalenko2020scalable} represents Wikipedia articles using Graph Convolutional Networks. Both works aim to recommend items (Wikipedia articles) to editors based on features of items and editors. There is no explicit feedback that would confirm an editor's interest in an article. Equally, there is very little information about editors beyond what they have edited so far \cite{moskalenko2020scalable}. Our system is similar in spirit to \cite{cosley2007suggestbot} and \cite{moskalenko2020scalable}, though  our neural approach incorporates additional item-based features and a mix of representations.

In community question-answering (CQA), several papers develop recommender systems to route users to questions they might be interested in answering, hence improving their engagement on the platform and reducing question answering time \cite{dror2011want,liu2017knowledge,sun2018coldroute,sun2019end}. \cite{dror2011want,sun2018coldroute} model recommendation as a classification problem, using different machine learning techniques. They implement a hybrid approach with content and collaborative knowledge to address the sparsity problem. \cite{liu2017knowledge,sun2019end} apply graph embedding techniques to tackle the same problem. Similarly, we mix item and relational representations with collaborative filtering to solve in a novel application context.   


Another related area is online crowdsourcing, where the aim is to allocate tasks published by a requester to an online crowd. \cite{kurup2017task,safran2018efficient} employ a probabilistic matrix factorization (PMF) to recommend suitable tasks to crowdworkers based on their previous activities, performance, and preferences. They handle the cold start problem by utilizing predefined categories (e.g., sentiment analysis, translation, image labeling) as additional features to improve the recommendation accuracy. 

\vspace{-3mm}
\subsection{Evaluating Recommender Systems}
\label{sec:RS-evaluation}
\noindent \emph{\textbf{Evaluation Methodologies.}}
To evaluate the performance of a recommender system, 
\textit{precision@k} and \textit{recall@k} are the most common metrics for top-k recommendations task with implicit data \cite{shani2011evaluating}. \textit{Precision@k} measures the relevance of the recommended items list, whereas \textit{recall@k} gives insight about how well the recommender is able to recall all the items the editor has rated positively (or interacted with) in the test set \cite{shani2011evaluating}. Moreover, there is a set of metrics that cares about where the relevant item appears in the recommended list. For instance, \textit{mean average precision(MAP)} and \textit{mean average recall (MAR)} are popular metric for search engines and are applied to the recommendation task~\cite{aggarwal2016recommender}. In these metrics, 
the relevant items are required to be placed as high on the recommendation list as possible \cite{shani2011evaluating}. There is also a family of metrics such as \textit{catalog coverage} and \textit{diversity} that pays special attention to editor experience \cite{aggarwal2016recommender}. The \textit{catalog coverage} evaluates whether the recommender algorithm can generate recommendations with a high proportion of items, and the \textit{diversity} evaluates how diverse the set of proposed recommendations within a single recommended list \cite{aggarwal2016recommender}.

\noindent \emph{\textbf{Recommender Datasets.}}
Recommender datasets are available for items such as movies (using MovieLens)\footnote{\url{https://grouplens.org/datasets/movielens/}} or products (using the Amazon dataset)\footnote{\url{https://jmcauley.ucsd.edu/data/amazon/}}. 
In other tasks, such benchmarks are not widely available. The systems discussed in Section \ref{sec:related-works} are mostly evaluated on custom-built datasets derived from the platforms' activity logs. This is also the case with our system, which to the best of our knowledge is the first of its kind for Wikidata. To encourage further research in this space, we hence provide two new datasets for the community to reuse, specified in Section \ref{sec:datasets}. 

\vspace{-3mm}
\section{Wikidata Recommender Datasets}
\label{sec:datasets}

\subsection{From Wikidata Dumps to Relational Tables}
\label{sec:datasetconstruction}

Wikidata dumps are readily available as JSON, RDF and XML\footnote{\url{https://dumps.wikimedia.org/wikidatawiki/}}. In our work we use the JSON and XML ones from July 1, 2019. The JSON dump contains all Wikidata pages without their edit history, which is available as XML. We parse the JSON data to extract all Wikidata items along with their corresponding identifiers, labels, descriptions, and statements. For the text data (i.e. label and description), we fetch only the English language version and discard data for other languages, because English is the best covered language in Wikidata \cite{kaffee2017glimpse}. Also, we ignore items’ metadata about aliases and sitelinks as we do not use them as features in our system. We transform the parsed data into CSV and import it into a PostgreSQL database (as a \emph{Wikidata-items-content} table); further processed the raw data containing the items’ statements (i.e. claims) in similar way (as a \emph{Wikidata-items-relations} PostgreSQL table). Next, we work on the XML dump to extract editing information for each individual editor. We only focus on edits that are performed on an item’s namespace and ignore all other non-item-related actions, as we do not model this information in our system. 
For each edit, the following information is kept: who carried out the edit, the item being edited, the timestamps of the edit, and the comment indicating the specific action executed by the editor on the item. They are stored in PostgreSQL (as a \emph{Wikidata-editors-edits} table). 

\subsection{Sampling and Cleaning}
\label{sec:sampling}
We randomly sample $14$ million editing activities performed by $221,353$ editors who are human registered (i.e. non-bots). We sort these activities in ascending order based on timestamps. We refer to this as the \textit{Wikidata-14M} dataset. 
We also want to test the system on a denser dataset that contains the editors who interacted with a larger number of items in their editing history and the items that received edits by a high number of editors. The \textit{Wikidata-14M} dataset is filtered by keeping editors who edited at least $200$ unique items during their tenure and items that have been edited by at least $5$ different editors. We refer to it as the \textit{Wikidata-Active-Editors} dataset. We then removed some outliers from both datasets - $2.14\%$ of editors in \textit{Wikidata-14M} and $5\%$ in \textit{Wikidata-Active-Editors} edited a very large number of items in a very short time relative to the size of their contributions. This is considered as the case for institutional accounts that publish their data via Wikidata \cite{turki2019wikidata}. We remove those accounts and their edits from the datasets, as they are not the main target editors for recommendations. 
We use both datasets to evaluate our system. The first one \textit{Wikidata-14M} depicts the actual diversity of editing activities in Wikidata with a mix of active and more casual editors. The second one \textit{Wikidata-Active-Editors} focuses on more active editors and items - we use it to understand the effects of data sparsity on recommender performance.



\begin{table}[t]
\caption{Number of items, editors and interactions in the two Datasets.}
\centering 
\label{table-1}
\begin{tabular}{|l|l|l|l|}
\hline
\textbf{Dataset} & \# editors & \# items & \# interactions  \\ \hline
\textbf{Wikidata-14M} & 221,353  & 4,881,720  & 14,045,523   \\ \hline
\textbf{Wikidata-Active-Editors} & 8,024 & 381,784 & 3,272,086 \\ \hline
\end{tabular}
\vspace{-3mm}
\end{table}

\vspace{-5mm}
\begin{table}[]
\caption{Statistics of the distribution of editing activities among editors and items.}
\centering 
\label{table-2}
\begin{tabular}{|l|l|l|l|l|l|l|}
\hline
\multirow{2}{*} & \multicolumn{3}{l|}{\textbf{Wikidata-14M}} & \multicolumn{3}{l|}{\textbf{Wikidata-Active-Editors}} \\ \cline{2-7} 
 & Median & Mean & Std & Median & Mean & Std \\ \hline
\# items/editor & 1 & 72 & 534 & 900 & 1,244 & 2,197 \\ \hline
\# edits/editor & 2 & 143 & 1,413 & 1,045 & 4,666 & 32,338 \\ \hline
\# editors/item & 1 & 2 & 3.7 & \_ & \_ & \_  \\ \hline
\end{tabular}
\end{table}

\vspace{-6mm}
\begin{figure}[!htb]
\captionsetup{justification=centering}
 \subfloat[Items vs. Editors]
{\minipage{0.32\textwidth}
  \includegraphics[height=4cm, width=\linewidth]{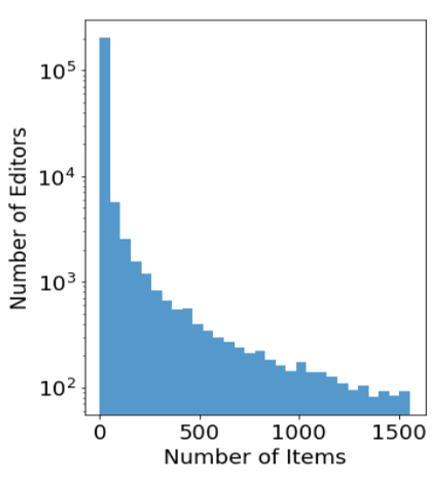}
  \label{2a}
\endminipage\hfill}
\subfloat[Editors vs. Items]
{\minipage{0.32\textwidth}%
  \includegraphics[height=4cm, width=\linewidth]{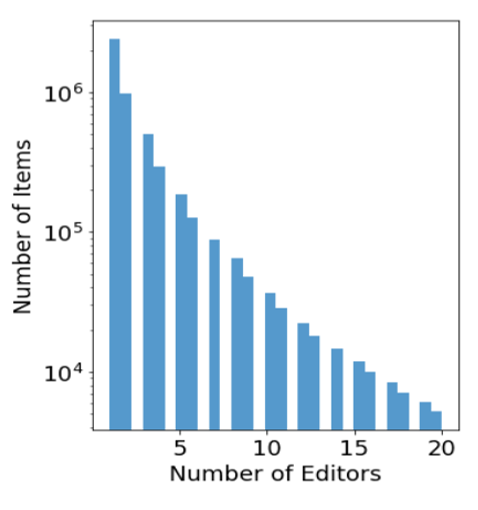}
  \label{2b}
\endminipage}
\subfloat[Items vs. Editors]
{\minipage{0.32\textwidth}%
  \includegraphics[height=4cm, width=\linewidth]{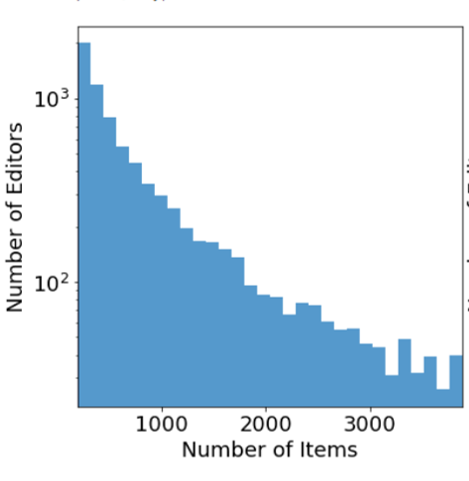}
  \label{2c}
\endminipage}
\caption{Editing Activities Distribution in \textit{Wikidata-14M} (a and b) and in \textit{Wikidata-Active-Editors} (c).}
\label{fig1}
\vspace{-4mm}
\end{figure}


\vspace{-3mm}
\subsection{Datasets Description}
\label{sec:dataset-content}

Table \ref{table-1} reports some key descriptive statistics of the two datasets, \ie number of editors, items and interactions. Figure \ref{2a}, \ref{2b} and \ref{2c} characterise the distribution of edits in the two datasets. For \textit{Wikidata-14M}, Figure \ref{2a} shows the number of items edited by each editor - most editors edited only a few items (<$1,000$); while fewer than $100$ editors (out of $221,353$) edited more than $1,500$ items. Table \ref{table-2} illustrates this skewed distribution: the median values are much smaller than the mean values, both per editor and per item. It is in line with observations from previous studies in Wikidata and other large-scale {community-driven knowledge engineering initiatives} \cite{kanza2018does,piscopo2018models,sarasua2019evolution}, which attest that a small core of contributors are responsible for a majority of the work. This also gives us an opportunity to increase retention in the long tail of editors (novice editors mostly) through recommendations. We also examine items that are edited by multiple editors, which is useful in collaborative filtering: Figure \ref{2b} shows that most items have been edited only by small number of editors; the highest number of editors per item is $20$. This implies the issue of cold-start for new items, we address it by mining item content and relation information for recommendation (see Section \ref{sec:model}).

We observe similar, though less pronounced effects for \textit{Wikidata-Active-Editors} (see Figure \ref{2c} and Table \ref{table-2}): the interaction data between editors and items are denser, e.g. the mean and median of the number of items per editor is $1,244$ and $900$, respectively,  showing a less skewed distribution. 
Notice we do not have the corresponding Figure~\ref{2b} (Editors vs. Items) for \textit{Wikidata-Active-Editors}, because all active editors are filtered and selected, making this statistic no longer meaningful. 

\section{Wikidata Recommender System} 
\label{sec:model}

\subsection{Problem Statement}
Our problem is defined as follows: given a set of $N$ editors, a set of $M$ items, and an interaction matrix $A^{N\times M} =\{a_{ij}\}$, where each matrix entry $a_{ij}$ is either {1 or 0} indicating whether editor $i$ has edited item $j$. The task is learning to estimate the preference scores of editors to items so as to rank and recommend new items to editors. Solving this problem is not straightforward: as discussed in Section \ref{sec:datasets}, there exists a high number of items but a low number of interactions between items and editors; many items have only a few edits. Standard approaches using collaborative filtering~\cite{schafer2007collaborative} rely primarily on the item-editor interaction data and do not perform well in our scenario. Intuitively, there are two ways to improve it: by including more information from either editors  (e.g.\ their interests or activities) or items (e.g.\ their content or relations). In this paper we focus on the latter, as for the former, there is very little descriptive information available for Wikidata editors, a known issue also present in Wikipedia~\cite{cosley2007suggestbot,moskalenko2020scalable}. Of course one could try to collect additional information about editors from their activities in Wikidata discussions or contributions to other WikiProjects. This however is not the scope of this paper as we decide to steer away from it to avoid the potential ethical implications.


\vspace{-3mm}
\subsection{Approach}
\label{sec:approach}
\begin{figure}[t]
    \includegraphics[width=0.9 \columnwidth]{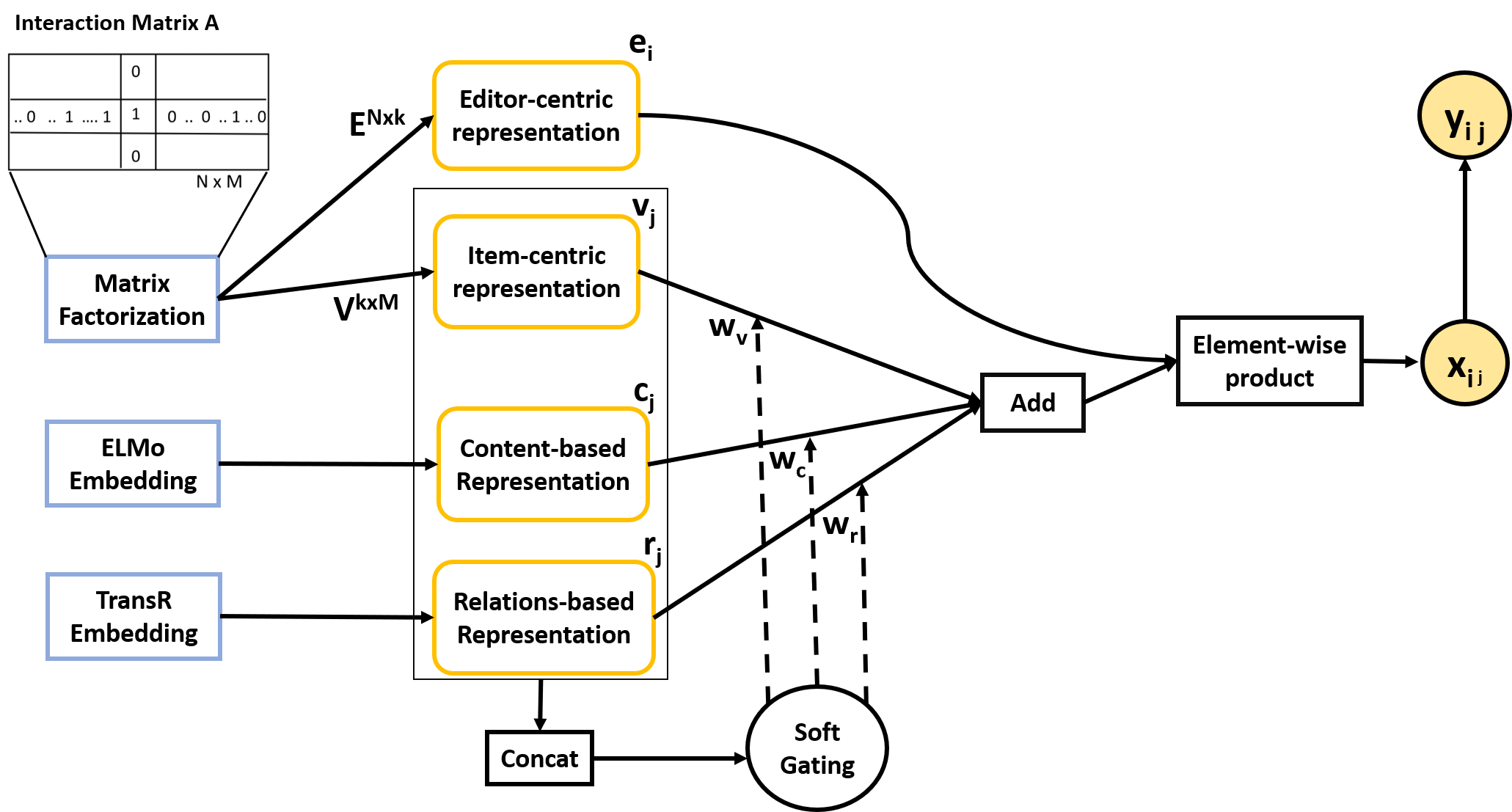}
    \centering 
    \caption{The architecture of \textit{WikidataRec}.}
    \label{fig:architecture}
     \vspace{-4mm}
\end{figure}

We introduce a Wikidata recommender system \textit{WikidataRec} which is a hybrid model combining item content and relation information with collaborative filtering~\cite{koren2009matrix} by means of mixture of experts (MoE)~\cite{jacobs1991adaptive} (Figure~\ref{fig:architecture}). The matrix factorization (MF) decomposes the item-editor interaction matrix $A$ into editor-centric and item-centric edit representations, denoted by $e_i$ and $v_j$ for editor $i$ and item $j$ respectively. Item content representation, denoted by $c_j$, is obtained by the sentence embedding model ELMo~\cite{peters2018deep}; and item relation representation, denoted by $r_j$, is obtained by the graph embedding model TransR~\cite{lin2015learning} building upon the Wikidata knowledge graph. To combine multiple item-based representations, we introduce a neural mixture of representations (NMoR) inspired by MoE, in which we utilize a soft-gating to assign different weights to each representation. More specifically, the network takes four inputs $e_i$, $v_j$, $c_j$ and $r_j$, where $v_j$, $c_j$ and $r_j$ are added with weights ($w_v$, $w_c$, $w_r$) produced from a soft-gating branch whose input is the concatenation of $v_j$, $c_j$ and $r_j$. The merged item representation is fed into an element-wise dot product layer together with $e_i$ to predict the preference score $x_{ij}$. At the training stage, $x_{ij}$ is optimized with the cross-entropy loss over the ground truth label in $A$; while at testing, it is used to recommend items to the editors. The whole process can be written as, 
\begin{equation}\label{eq:FinalModel}
     {x_{ij}} =  e_i (w_v\cdot v_j + w_c \cdot c_j+ w_r \cdot r_j)^\intercal.
\end{equation}
Notice item and editor representations are used in the network as fixed representations (i.e., they are not updated during the training process). They are learned in advance by MF, ELMo, and TransR, respectively. We decide not to jointly tune these representations for efficiency and simplicity reasons, as it is not lightweight to integrate all the models in a whole. We will work on joint learning in our future work.

Below we first specify the generation of $e_i$, $v_j$, $c_j$ and $r_j$ and then introduce the neural mixture of representations for $w_v$, $w_c$ and $w_r$. 

\subsection{Editor-centric and Item-centric Edit Representations} 
\label{sec:latent}
Editing activities are summarized in the interaction matrix $A$. Inspired by~\cite{koren2009matrix}, we use matrix factorization to decompose $A$ into the editor’s latent representation matrix $E^{N \times Z}$ and the item’s latent representation matrix $V^{M\times Z}$. This is achieved by approximating the target matrix $A$ via the matrix product of two low-rank matrices $E$ and $V$:

\begin{equation} \label{eq:latent}
     \widehat{A} = EV^{\intercal}  
\end{equation}

\noindent Each row $e_i$ in $E$ can be seen as a latent representation
of editor $i$ in terms of its edits. Similarly, each row $v_j$ of $V$ describes an item $j$ with respect to its edits by different editors. Thus, the prediction formula from Eqn(\ref{eq:latent}) can also be written as: 

\begin{equation} \label{eq:latent_element}
     \widehat{a_{ij}} = e_iv_j^{\intercal}  
\end{equation}

\noindent In order to single out $E$, $V$ from $A$, we utilize Bayesian Personalized Ranking (BPR)~\cite{rendle2012bpr} to optimize $\widehat{A}$ over it. If we look at Eqn(\ref{eq:FinalModel}), $\widehat{a_{ij}}$($e_iv_j^{\intercal}$) is its first term.

\vspace{-3mm}
\subsection{Item Content Representation} 
\label{sec:content}

To generate item content representation $c_j$, we need an embedding model that can learn both semantic and syntactic representations from text. There are various embedding models that learn item representations from words, sentences, or paragraphs. Among them, word embedding models such as Word2Vec~\cite{mikolov2013efficient}, GloVe \cite{pennington2014glove} and FastText~\cite{joulin2016bag} are widely used for many recommendation tasks~\cite{esmeli2020using}, but they have limitations: the order of words is ignored, which leads to the loss of the syntactic and semantic meaning in sentences~\cite{wang2019evaluating}. We therefore considered two state-of-the-art sentence embedding models instead, specifically Embeddings from Language Models (ELMo) and Sentence-BERT. ELMo~\cite{peters2018deep} employs a bi-directional deep LSTM network which takes an entire sentence as input, assigns representation to each word in the sentence, then takes the average of the vectors of words to output a $Z$-dimensional vector for the input sentence. Sentence-BERT \cite{reimers2019sentence} relies on a bi-directional Transformer network \cite{reimers2019sentence} which learns from fix-sized semantically meaningful sentences. We preferred ELMo to Sentence-BERT to generate $c_j$ because of the high computational costs of the latter, particularly on large corpora \cite{polignano2020contextualized}. Each item in Wikidata has a very short description which naturally serves as input sentence to ELMo. The implementation details are in Section~\ref{sec:setup}.

\vspace{-3mm}
\subsection{Item Relational Representation} 
\label{sec:relation}
Item relational reprentations are built upon the Wikidata graph over items. We model item relations as a directed labeled graph, where it has a set of nodes to represent items, a set of edges (unweighted) to represent relations between items, and labels to capture the type of relations (see Section \ref{sec:data_model}). Over different graph embedding models~\cite{xiao2015transa,bordes2013translating,lin2015learning}, we adopt the TransR~\cite{lin2015learning}, a representative approach for heterogeneous graphs, to learn low-dimensional ($Z$ in this work) relational embedding of items, $r_j$. It builds entity and relation embeddings by regarding a relation as a translation from head entity to tail entity~\cite{lin2015learning}. Different from other graph embedding methods\cite{xiao2015transa,bordes2013translating} which normally assume the embedding of entities and relations within the same space, TransR represents entities and relations in distinct spaces and projects entities from entity space to relation space via a projection matrix. For this reason, it is selected to model the heterogeneity of items and their relations in the Wikidata graph. Items with similar relations have similar embedding in TransR. The implementation details are in Sec.~\ref{sec:setup}. 

\vspace{-3mm}
\subsection{Neural Mixture of Representations} 
\label{sec:NMoE}
Having $e_i$, $v_j$, $c_j$, and $r_j$ ready, we present our neural mixture of representations (NMoR) to combine them for Wikidata recommendation. $e_i$, $v_j$, $c_j$, and $r_j$ are fed into NMoR in parallel as fixed representations, where they are not updated during training but learned in advance (as noted in Section \ref{sec:approach}). Item representations $v_j$, $c_j$ and $r_j$ are added with weights $w_v$, $w_c$ and $w_r$ generated by an additional soft-gating branch (see Figure~\ref{fig:architecture}). Soft-gating is used to distinguish from hard-gating in MoE. In hard-gating, weights are either 0 or 1 for each expert. Soft-gating can be seen as probabilities combining different experts. The motivation for assigning different weights to different item-based representations is to control their respective contribution to the final prediction $x_{ij}$ (Eqn.~\ref{eq:FinalModel}). These weights are also $Z$-dimensional vectors as with $v_j$, $c_j$, and $r_j$ such that they are tailored to every dimension of feature representations. The soft-gating branch takes the concatenation of $v_j$, $c_j$ and $r_j$ of size 1 $\times$ Z $\times$ 3 (corresponding to rows, columns, and channels of the tensor) as input; 
three 1D convolutional layers (kernel size 1 x 1, filters $1024$) followed by ReLu are applied to process the tensor along its channels; the number of their output channel remains $3$. The output of the last convolutional layer is thus of size 1 $\times$ Z $\times$ 3, where each channel corresponds to the weight for certain item-based representation ($w_v$, $w_c$ or $w_r$). At every column of the tensor, it is softmaxed over channels such that the three values at each column are summed to $1$. The merged item representation $w_v\cdot v_j + w_v \cdot c_j+ w_r \cdot r_j$ is passed through the element-wise dot product layer with $e_i$, their similarity value $x_{ij}$ is hence computed. To optimize $x_{ij}$, we adopt the binary cross-entropy loss,
\begin{equation} \label{eq:loss}
     L = - (y_{ij} \log(\sigma(x_{ij})) + (1-y_{ij})\log(1- \sigma(x_{ij}))), 
\end{equation}
where $x_{ij}$ is transformed into a probability using the sigmoid activation function $\sigma (\cdot)$; $y_{ij}$ is the ground truth label of either 0 or 1 in $A$ indicating whether interaction between the item $j$ and editor $i$ has been observed (\ie positive instances) or not (\ie negative instances). Since $A$ is sparse, we randomly select negative instances to make its number the same to that of positive instances. $x_{ij}$ indicates how likely item $j$ is relevant to editor $i$, we can use it to recommend items to editors at testing.

Traditional weight tuning approach like line search is only suitable for optimizing a limited number of weights, for instance, one parameter for each item-based representation. The proposed NMoR optimizes weights in a neural network which enables a refined weight steering on every dimension of the item-based representation. 

\section{Experiments}
\label{sec:experiments}

\subsection{Experimental Setup and Implementation Details}\label{sec:setup}
We run experiments on the two datasets \textit{Wikidata-14M} and \textit{Wikidata-Active-Editors} introduced in Section \ref{sec:datasets}. Both datasets are structured in an editor-item-edits format. For each dataset, similar to~\cite{he2017neural,he2016fast}, we follow the hold-out strategy to select $80\%$ items associated with each editor to constitute the training set and use the remaining $20\%$ as the test set. Within the training set, we set aside $10\%$ as validation data for hyper-parameter tuning. In \textit{Wikidata-14M} dataset, $25\%$ of the editors edited only between 2 to 6 different items during their editing tenure on the selected dates; these are cold-start editors. Therefore, we exclude these editors from the training set and include them only in the test set. We do this because the editing data for this type of editor are rather too sparse to be informative during training. However, we use this data in the test set, as we want to evaluate the performance of \textit{WikidataRec} on cold-start editors. $Z$ for feature representations is set to $1024$. 

\noindent \emph{\textbf{NMoR:}} To learn NMoR, the batch size is set to $32$. The model is learned with Adam optimizer with a learning rate of $0.001$ for 100 epochs. 

\noindent \emph{\textbf{ELMo:}} To generate the item content representations, we start from the \textit{Wikidata-items-content} table in our datasets, and then follow the following steps using Python's Spacy:
1) tokenize and normalize the content (label and description) of each item; 2) remove stopwords, punctuation and single-letter words; 3) use part-of-speech tagging and retained only nouns and adjectives. We pass the resulting corpus as input to the ELMo model. ELMo is pre-trained on the $1$ Billion Word Benchmark that contains about $800M$ tokens of general-purpose data from WMT 2011\cite{chelba2013one}. This is very comprehensive such that fine-tuning on Wikidata is no longer necessary. Following~\cite{hassan2019bert}, we simply forward each Wikidata item descriptor to ELMo to obtain the item-content representation. Notice ELMo produces multi-scale output which we average them. 

\noindent \emph{\textbf{TransR:}} To generate the item relational representations, we first build the labeled directed graph $G(V,E,P)$ from the \textit{Wikidata-items-relations} (Sec.~\ref{sec:datasetconstruction}), where $V$ signifies Wikidata items (the head of the triplet), $E$ inter-item relations (the tail of the triplet), and $P$ relations' types (the relation of the triplet). We use the graph as input to train the TransR model. The original triplets in the graph are the positive instances. We sample a few of them to replace either their heads or tails with wrong components to create negative triplets. TransR is trained with both positive and negative triplets using binary cross-entropy loss. It is optimized with the SGD optimizer for $10$ epochs; the batch size is $128$ and learning rate is $0.01$. 

\vspace{-3mm}
\subsection{Evaluation Protocols}

For evaluation, in order to simulate the practical recommendation scenario in Wikidata, we follow the spirit in~\cite{moskalenko2020scalable} to carry out the test: for each editor $i$ and the item-editor interaction matrix $A$ in the test set,  we 1) split the items of editor $i$ into half-half where the editor-centric edit representation $e_i$ along with the representations of item-centric edit, content, and relations $v_j$, $c_j$ and $r_j$ are obtained by excluding the second-half items in $A$ in our model; 2) exclude the editor $i$ from $A$ to generate the representations of item-centric edits, item-content and item relations for the second half of items in our model; 3) randomly select $200$ negative items that were not edited by editor $i$, where we can obtain their corresponding sub-matrix of $A$ and obtain their item-based representations; 4) feed ($e_i$, $v_j$, $c_j$ and $r_j$) obtained from step 1 to our NMoR along with the item-based representations obtained from steps 2 and 3 to obtain their ranking scores; 

The evaluation criteria is to measure how well \emph{WikidataRec} ranks the correct items against the negative items for a given editor. Therefore, we employ \emph{Precision@k}, \emph{Recall@k}, and \emph{MAR@k} (see Section \ref{sec:RS-evaluation}).  
We are also interested in the diversity of the recommended items, as editing a different set of items is a noted behavior in Wikidata community \cite{sarasua2019evolution,piscopo2017makes}. Having a diverse recommendation would allow the editors to discover a wide range of items for the editing.


\begin{table}[t]
\caption{Precision@k and Recall@k results comparison between our model and state-of-the-art for \emph{Wikidata-14M}.}
\centering
	\setlength{\tabcolsep}{5pt}
\label{Table-Performance-Comparison}
\begin{tabular}{llllll}
\toprule
{\color[HTML]{000000} } & \multicolumn{2}{c|}{{\color[HTML]{000000} \textbf{Precision @k}}} & \multicolumn{3}{c}{{\color[HTML]{000000} \textbf{Recall @k}}} \\ 
\cline{2-6} 
\multirow{2}{*}{{\color[HTML]{000000} \textbf{}}} & \multicolumn{1}{c|}{{\color[HTML]{000000} 5}} & \multicolumn{1}{c|}{{\color[HTML]{000000} 10}} & \multicolumn{1}{c|}{{\color[HTML]{000000} 50}} & \multicolumn{1}{c|}{{\color[HTML]{000000} 100}} & \multicolumn{1}{c}{{\color[HTML]{000000} 200}} \\
\midrule
{\color[HTML]{000000} GMF} & {\color[HTML]{000000} 0.096} & {\color[HTML]{000000} 0.071} & {\color[HTML]{000000} 0.135} & {\color[HTML]{000000} 0.183} & {\color[HTML]{000000} 0.274} \\ 
{\color[HTML]{000000} BPR-MF} & {\color[HTML]{000000} 0.050} & {\color[HTML]{000000} 0.043} & {\color[HTML]{000000} 0.093} & {\color[HTML]{000000} 0.135} & {\color[HTML]{000000} 0.190} \\ 
{\color[HTML]{000000} eALS} & {\color[HTML]{000000} 0.048} & {\color[HTML]{000000} 0.027} & {\color[HTML]{000000} 0.087} & {\color[HTML]{000000} 0.112} & {\color[HTML]{000000} 0.154} \\ 
{\color[HTML]{000000} YouTube-DNN} & {\color[HTML]{000000} 0.105} & {\color[HTML]{000000} 0.082} & {\color[HTML]{000000} 0.142} & {\color[HTML]{000000} 0.204} & {\color[HTML]{000000} 0.305} \\ 
{\color[HTML]{000000} WikidataRec} & {\color[HTML]{000000} \textbf{0.120}} & {\color[HTML]{000000} \textbf{0.113}} & {\color[HTML]{000000} \textbf{0.215}} & {\color[HTML]{000000} \textbf{0.243}} & {\color[HTML]{000000} \textbf{0.337}} \\ \bottomrule
\end{tabular}
\end{table}

\subsection{Results on Wikidata-14M}
\noindent \textbf{Comparison to state of the art.} We compare \emph{WikidataRec} against the following methods on \emph{Wikidata-14M}: \emph{\textbf{BPR-MF}} \cite{rendle2012bpr} is a representative collaborative filtering model that uses matrix factorization (MF) as the underlying predictor and is optimized with a pairwise ranking loss. It is suitable for recommendation scenarios with no explicit editor feedback and personalised ranked recommendation results.  \emph{\textbf{eALS}} \cite{he2016fast} is also a MF-based method that is optimized with square loss. It treats all unobserved interactions as negative instances and weights them non-uniformly by the item's popularity.
 \emph{\textbf{GMF}} \cite{he2017neural} is a neural network-based collaborative filtering which implements MF with cross-entropy loss. It embeds each item and editor in the network and computes their element-wise dot product to predict the relevance score.
\emph{\textbf{YouTube-DNN}} \cite{covington2016deep} is a neural recommender for YouTube videos, using deep candidate generation and ranking networks. It uses a hybrid of users' activities and content information of users and items and directly learns their low dimensional representations. In this paper, we adapt YouTube-DNN with our item content and relational representations. 

Table \ref{Table-Performance-Comparison} shows the results: \emph{WikidataRec} achieves the best performance over all with both accurate and diverse recommendations. First, it outperforms the collaborative-filtering (CF) methods GMF, eALS and BPR-MF by a large margin. CF methods only rely on the item-editor interactions data without taking into account the information of items themselves, whereas adding additional item content and relation information significantly improves the recommendation performance in our model (see Table~\ref{Table-Ablation-study-on-features}). Our model also beats Youtube DNN. Youtube-DNN learns item's content and relational embedding in a neural network with random initialization while we employ additional state-of-the-art embedding models (i.e., ELMo and TransR) to generate them separately which yields superior performance.


\noindent \textbf{Ablation Study. } We ablate different components of \emph{WikidataRec} on the \emph{Wikidata-14M} to understand the effect of each one.

\noindent \textbf{\emph{Item contents and relations.}} To justify the importance of item content and relations in \emph{WikidataRec}, we start with the original collaborative filtering (CF) technique, BPR-MF, and gradually add item content and relational representations using NMoR. The results are in Table  \ref{Table-Ablation-study-on-features} which show that values in every metric are increased by considering item content and relations. Particularly, item content representations provide rich information in terms of words and semantic meanings in items, which contribute more in performance increase; item relational representations in contrast contribute less. We can note each feature's contributions from the amount of increase when each of them is added. We suggest the reason as these relations are learned from the Wikidata graph, which is unevenly and sparingly connected, as discussed in \cite{piscopo2018models}. 

\noindent \textbf{\emph{Neural mixture of representations.}} To ablate the proposed NMoR, Table \ref{Table-Ablation-study-on-features} further illustrates the results of \emph{WikidataRec} with and without using it. For the latter,  the item-based representations are added with no weights. WikidataRec with NMoR performs significantly better than WikidataRec without NMoR.

\begin{table}[t]
	\setlength{\tabcolsep}{4pt}
\caption{Ablation study of \emph{WikidataRec} on \emph{Wikidata-14M}.}
\centering
\label{Table-Ablation-study-on-features}
\begin{tabular}{lllllllllll}
\toprule
{\color[HTML]{000000} } & \multicolumn{2}{c|}{{\color[HTML]{000000} \textbf{Precision @k}}} & \multicolumn{3}{c|}{{\color[HTML]{000000} \textbf{Recall @k}}} & \multicolumn{2}{c|}{{\color[HTML]{000000} \textbf{MAR@k}}} & \multicolumn{1}{c}{{\color[HTML]{000000} \textbf{Diversity}}} \\ \cline{2-9} 
\multirow{2}{*}{{\color[HTML]{000000} \textbf{}}} & \multicolumn{1}{c|}{{\color[HTML]{000000} 5}} & \multicolumn{1}{c|}{{\color[HTML]{000000} 10}} & \multicolumn{1}{c|}{{\color[HTML]{000000} 50}} & \multicolumn{1}{c|}{{\color[HTML]{000000} 100}} & \multicolumn{1}{c|}{{\color[HTML]{000000} 200}} & \multicolumn{1}{c|}{{\color[HTML]{000000} 5}} & \multicolumn{1}{c|}{{\color[HTML]{000000} 10}} & \multicolumn{1}{c}{{\color[HTML]{000000} 10}} \\ \hline
{\color[HTML]{000000} CF (BPR-MF)} & {\color[HTML]{000000} 0.050} & {\color[HTML]{000000} 0.043} & {\color[HTML]{000000} 0.093} & {\color[HTML]{000000} 0.135} & {\color[HTML]{000000} 0.190} & {\color[HTML]{000000} 0.069} & {\color[HTML]{000000} 0.107} & {\color[HTML]{000000} 0.252} \\ 
{\color[HTML]{000000} \begin{tabular}[c]{@{}l@{}}+ item content\end{tabular}} & {\color[HTML]{000000} \begin{tabular}[c]{@{}l@{}}0.104 \end{tabular}} & {\color[HTML]{000000} \begin{tabular}[c]{@{}l@{}}0.086 \end{tabular}} & {\color[HTML]{000000} \begin{tabular}[c]{@{}l@{}}0.189 \end{tabular}} & {\color[HTML]{000000} \begin{tabular}[c]{@{}l@{}}0.199 \end{tabular}} & {\color[HTML]{000000} \begin{tabular}[c]{@{}l@{}}0.287 \end{tabular}} & {\color[HTML]{000000} \begin{tabular}[c]{@{}l@{}}0.108 \end{tabular}} & {\color[HTML]{000000} \begin{tabular}[c]{@{}l@{}}0.170 \end{tabular}} & {\color[HTML]{000000} \begin{tabular}[c]{@{}l@{}}0.502 \end{tabular}} \\ 
{\color[HTML]{000000} \begin{tabular}[c]{@{}l@{}}+ item relations \end{tabular}} & {\color[HTML]{000000} \begin{tabular}[c]{@{}l@{}}\textbf{0.120}\end{tabular}} & {\color[HTML]{000000} \begin{tabular}[c]{@{}l@{}}\textbf{0.113} \end{tabular}} & {\color[HTML]{000000} \begin{tabular}[c]{@{}l@{}}\textbf{0.215}\end{tabular}} & {\color[HTML]{000000} \begin{tabular}[c]{@{}l@{}}\textbf{0.243}\end{tabular}} & {\color[HTML]{000000} \begin{tabular}[c]{@{}l@{}}\textbf{0.337}\end{tabular}} & {\color[HTML]{000000} \begin{tabular}[c]{@{}l@{}}\textbf{0.133}\end{tabular}} & {\color[HTML]{000000} \begin{tabular}[c]{@{}l@{}}\textbf{0.224}\end{tabular}} & {\color[HTML]{000000} \begin{tabular}[c]{@{}l@{}}\textbf{0.567}\end{tabular}} \\

\midrule
{\color[HTML]{000000} \begin{tabular}[c]{@{}l@{}}WikidataRec w/o NMoR\end{tabular}} & {\color[HTML]{000000} \begin{tabular}[c]{@{}l@{}}0.085\end{tabular}} & {\color[HTML]{000000} \begin{tabular}[c]{@{}l@{}}0.072\end{tabular}} & {\color[HTML]{000000} \begin{tabular}[c]{@{}l@{}}0.165\end{tabular}} & {\color[HTML]{000000} \begin{tabular}[c]{@{}l@{}}0.209\end{tabular}} & {\color[HTML]{000000} \begin{tabular}[c]{@{}l@{}}0.284\end{tabular}} & {\color[HTML]{000000} \begin{tabular}[c]{@{}l@{}}0.090\end{tabular}} & {\color[HTML]{000000} \begin{tabular}[c]{@{}l@{}}0.191\end{tabular}} & {\color[HTML]{000000} \begin{tabular}[c]{@{}l@{}}0.521\end{tabular}} \\
{\color[HTML]{000000} \begin{tabular}[c]{@{}l@{}}WikidataRec w/ NMoR\end{tabular}} & {\color[HTML]{000000} \begin{tabular}[c]{@{}l@{}}\textbf{0.120}\end{tabular}} & {\color[HTML]{000000} \begin{tabular}[c]{@{}l@{}}\textbf{0.113}\end{tabular}} & {\color[HTML]{000000} \begin{tabular}[c]{@{}l@{}}\textbf{0.215}\end{tabular}} & {\color[HTML]{000000} \begin{tabular}[c]{@{}l@{}}\textbf{0.243}\end{tabular}} & {\color[HTML]{000000} \begin{tabular}[c]{@{}l@{}}\textbf{0.337}\end{tabular}} & {\color[HTML]{000000} \begin{tabular}[c]{@{}l@{}}\textbf{0.133}\end{tabular}} & {\color[HTML]{000000} \begin{tabular}[c]{@{}l@{}}\textbf{0.224}\end{tabular}} & {\color[HTML]{000000} \begin{tabular}[c]{@{}l@{}}\textbf{0.567}\end{tabular}}\\
 \bottomrule
\end{tabular}
  \vspace{-4mm}
\end{table}

\begin{table}[t]
	\setlength{\tabcolsep}{5pt}
\caption{Ablation study of embedding methods for item content on \emph{Wikidata-14M}. }
\centering
\label{Table-embeddings-methods}
\begin{tabular}{lllll}
\toprule
{\color[HTML]{000000} }                                     & {\color[HTML]{000000} \textbf{Embedding Model}} & {\color[HTML]{000000} \textbf{Precision@5}} & {\color[HTML]{000000} \textbf{Recall@50}} & {\color[HTML]{000000} \textbf{Recall@100}} 
\\ 
\midrule
{\color[HTML]{000000} }                                     & {\color[HTML]{000000} Word2Vec}        & {\color[HTML]{000000} 0.060}      & {\color[HTML]{000000} 0.123}    & {\color[HTML]{000000} 0.158}     \\ 
 \multirow{2}{*}{{\color[HTML]{000000} BPR-MF + }}& {\color[HTML]{000000} FastText}        & {\color[HTML]{000000} 0.029}      & {\color[HTML]{000000} 0.069}    & {\color[HTML]{000000} 0.115}     \\ 
{\color[HTML]{000000} }                                     & {\color[HTML]{000000} ELMo}            & {\color[HTML]{000000} \textbf{0.104}}      & {\color[HTML]{000000} \textbf{0.189}}    & {\color[HTML]{000000} \textbf{0.199}}     \\ \bottomrule
\end{tabular}
  \vspace{-4mm}
\end{table}

\noindent \textbf{\emph{Embedding models for item content.}} 
We compare the ELMo model with two text embedding models, Word2Vec~\cite{mikolov2013efficient} and FastText~\cite{joulin2016bag} to generate item content representations. Word2Vec and FastText are lightweight models which we train them from scratch using our Wikidata. The results are in Table \ref{Table-embeddings-methods}: ELMo outperforms Word2Vec and FastText clearly. ELMo employs the deep bi-directional Language Model (biLM), which provides a very rich representation about the word tokens and captures the semantic and syntactic meaning of words. This is not the case in the Word2Vec and FastText who utilize shallow neural networks.

\vspace{-7pt}
\subsection{Results on Wikidata-Active-Editors}

We ran another ablation study of our model on the second dataset, \emph{Wikidata-Active-Editors}, introduced in Section~\ref{sec:dataset-content}. \emph{Wikidata-Active-Editors} is a subset of \emph{Wikidata-14M}, focusing on active editors and frequently edited items. The results are in Table \ref{Table-Wikidata-Active-Editor-Results}, in which we note improved results on \emph{WikidataRec} by adding item content and relations. In particular, content-based information contributed most towards the performance of the model. On the other hand, the contribution of relational information is less than that of content information. Furthermore, the results show that \emph{WikidataRec} with NMoR performs better than \emph{WikidataRec} without NMoR.

\begin{figure}[t]
    \includegraphics[width=0.5\columnwidth]{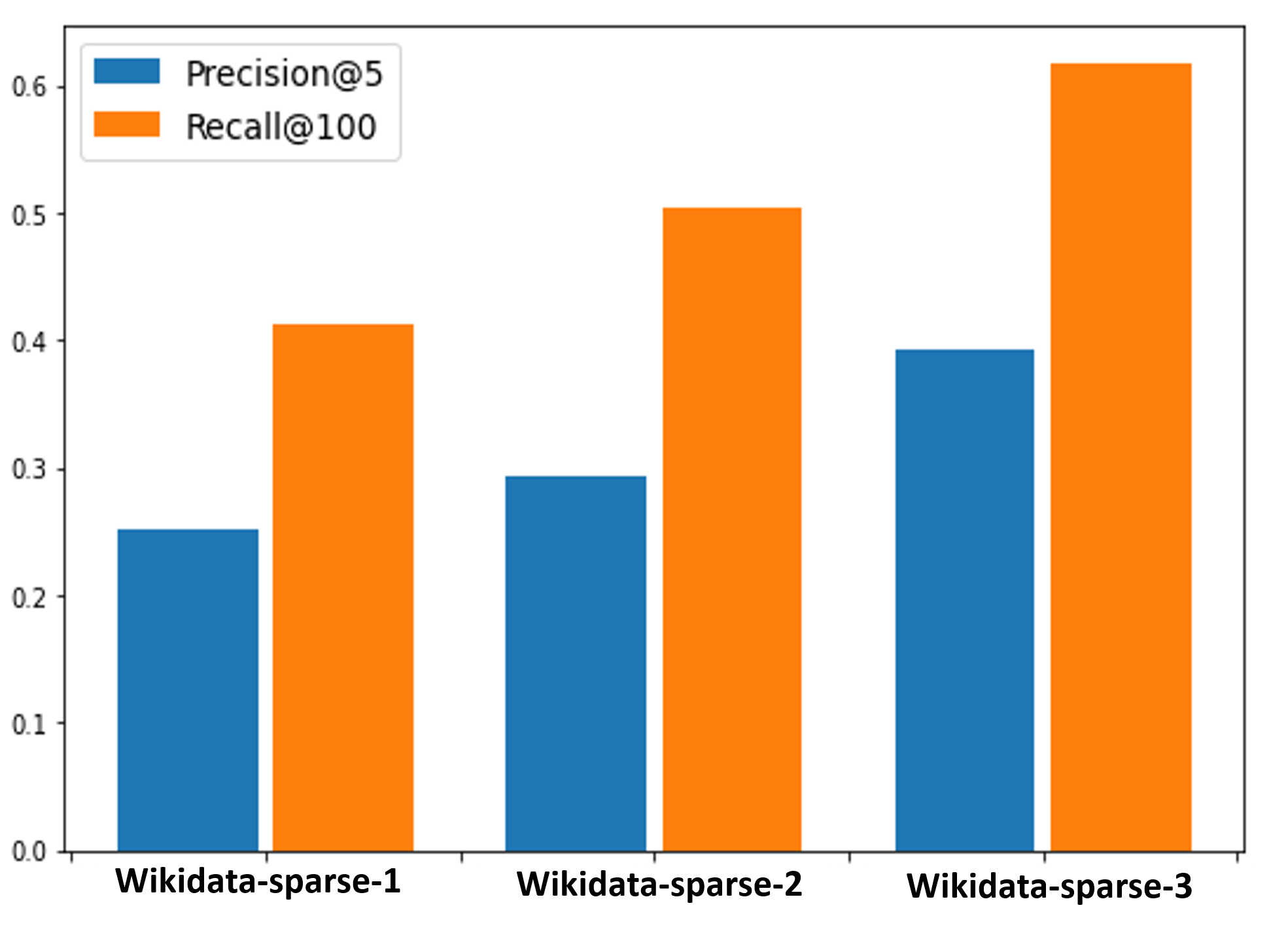}
    \centering 
    \caption{\textit{WikidataRec} with various sparsity levels.}
    \label{fig:desity-experiment}
   \vspace{-5mm}
\end{figure}

\noindent \textbf{\emph{Dataset sparsity.}} We further study the sparsity of the dataset in terms of its item-editor interaction data. Wikidata is very sparse in nature: a small number of interactions between items and editors are observed. \emph{Wikidata-Active-Editors} is denser than \emph{Wikidata-14M}, but its sparsity is still $99.90\%$, which means only 1\% of interactions are observed over all the possible connections from every item to every editor in the dataset. To study the influence of the dataset sparsity, we extract three sub-datasets from \emph{Wikidata-Active-Editors} with different sparsity yet roughly the same size (about 6000 editors and 10,000 items): \textit{Wikidata-sparse-1} with $99.81\%$ sparsity, \textit{Wikidata-sparse-2} with $99.68\%$ sparsity, and \textit{Wikidata-sparse-3} with $99.27\%$. The editors in each dataset have edited more than $200$ different items, and each item has edited by more than $16$ different editors. 
Figure \ref{fig:desity-experiment} show the Precision@5 and Recall@10 on the three subsets. It shows that when the sparsity decreases (from $99.81\%$ to $99.27\%$, data gets denser), our model performance increases. This again verifies the conclusion in~\cite{adomavicius2012impact}. 

\subsection{Discussion and analysis of results on both datasets}
Comparing the results of the two datasets (Table \ref{Table-Ablation-study-on-features} and Table \ref{Table-Wikidata-Active-Editor-Results}), we summarize that: 1) adding item content and relations with NMoR improves the performance on both datasets; 2) WikidataRec performs better on the \emph{Wikidata-Active-Editors} than on the \emph{Wikidata-14M}; the former is with denser editing data, so having less sparse data is likely to improve the results. This finding is consistent with the conclusion of \cite{adomavicius2012impact}.

\begin{table}[t]
	\setlength{\tabcolsep}{4pt}
\caption{Ablation study of \emph{WikidataRec} on \emph{Wikidata-Active-Editor}.}
\centering
\label{Table-Wikidata-Active-Editor-Results}
\begin{tabular}{lllllllllll}
\toprule
{\color[HTML]{000000} } & \multicolumn{2}{c|}{{\color[HTML]{000000} \textbf{Precision @k}}} & \multicolumn{3}{c|}{{\color[HTML]{000000} \textbf{Recall @k}}} & \multicolumn{2}{c|}{{\color[HTML]{000000} \textbf{MAR@k}}} & \multicolumn{1}{c}{{\color[HTML]{000000} \textbf{Diversity}}} \\ \cline{2-9} 
\multirow{2}{*}{{\color[HTML]{000000} \textbf{}}} & \multicolumn{1}{c|}{{\color[HTML]{000000} 5}} & \multicolumn{1}{c|}{{\color[HTML]{000000} 10}} & \multicolumn{1}{c|}{{\color[HTML]{000000} 50}} & \multicolumn{1}{c|}{{\color[HTML]{000000} 100}} & \multicolumn{1}{c|}{{\color[HTML]{000000} 200}} & \multicolumn{1}{c|}{{\color[HTML]{000000} 5}} & \multicolumn{1}{c|}{{\color[HTML]{000000} 10}} & \multicolumn{1}{c}{{\color[HTML]{000000} 10}} \\ \hline
{\color[HTML]{000000} CF (BPR-MF)} & {\color[HTML]{000000} 0.079} & {\color[HTML]{000000} 0.055} & {\color[HTML]{000000} 0.139} & {\color[HTML]{000000} 0.260} & {\color[HTML]{000000} 0.236} & {\color[HTML]{000000} 0.115} & {\color[HTML]{000000} 0.193} & {\color[HTML]{000000} 0.353} \\ 
{\color[HTML]{000000} \begin{tabular}[c]{@{}l@{}}+ item content\end{tabular}} & {\color[HTML]{000000} \begin{tabular}[c]{@{}l@{}}0.143\end{tabular}} & {\color[HTML]{000000} \begin{tabular}[c]{@{}l@{}}0.109\end{tabular}} & {\color[HTML]{000000} \begin{tabular}[c]{@{}l@{}}0.253\end{tabular}} & {\color[HTML]{000000} \begin{tabular}[c]{@{}l@{}}0.302\end{tabular}} & {\color[HTML]{000000} \begin{tabular}[c]{@{}l@{}}0.349\end{tabular}} & {\color[HTML]{000000} \begin{tabular}[c]{@{}l@{}}0.157\end{tabular}} & {\color[HTML]{000000} \begin{tabular}[c]{@{}l@{}}0.251\end{tabular}} & {\color[HTML]{000000} \begin{tabular}[c]{@{}l@{}}0.565\end{tabular}} \\ 
{\color[HTML]{000000} \begin{tabular}[c]{@{}l@{}}+ item relations \end{tabular}} & {\color[HTML]{000000} \begin{tabular}[c]{@{}l@{}}\textbf{0.164}\end{tabular}} & {\color[HTML]{000000} \begin{tabular}[c]{@{}l@{}}\textbf{0.131} \end{tabular}} & {\color[HTML]{000000} \begin{tabular}[c]{@{}l@{}}\textbf{0.289}\end{tabular}} & {\color[HTML]{000000} \begin{tabular}[c]{@{}l@{}}\textbf{0.342}\end{tabular}} & {\color[HTML]{000000} \begin{tabular}[c]{@{}l@{}}\textbf{0.391}\end{tabular}} & {\color[HTML]{000000} \begin{tabular}[c]{@{}l@{}}\textbf{0.179}\end{tabular}} & {\color[HTML]{000000} \begin{tabular}[c]{@{}l@{}}\textbf{0.297}\end{tabular}} & {\color[HTML]{000000} \begin{tabular}[c]{@{}l@{}}\textbf{0.596}\end{tabular}} \\

\midrule
{\color[HTML]{000000} \begin{tabular}[c]{@{}l@{}}WikidataRec w/o NMoR\end{tabular}} & {\color[HTML]{000000} \begin{tabular}[c]{@{}l@{}}0.132\end{tabular}} & {\color[HTML]{000000} \begin{tabular}[c]{@{}l@{}}0.073\end{tabular}} & {\color[HTML]{000000} \begin{tabular}[c]{@{}l@{}}0.196\end{tabular}} & {\color[HTML]{000000} \begin{tabular}[c]{@{}l@{}}0.295\end{tabular}} & {\color[HTML]{000000} \begin{tabular}[c]{@{}l@{}}0.323\end{tabular}} & {\color[HTML]{000000} \begin{tabular}[c]{@{}l@{}}0.134\end{tabular}} & {\color[HTML]{000000} \begin{tabular}[c]{@{}l@{}}0.245\end{tabular}} & {\color[HTML]{000000} \begin{tabular}[c]{@{}l@{}}0.553\end{tabular}} \\
{\color[HTML]{000000} \begin{tabular}[c]{@{}l@{}}WikidataRec w/ NMoR\end{tabular}} & {\color[HTML]{000000} \begin{tabular}[c]{@{}l@{}}\textbf{0.164}\end{tabular}} & {\color[HTML]{000000} \begin{tabular}[c]{@{}l@{}}\textbf{0.131}\end{tabular}} & {\color[HTML]{000000} \begin{tabular}[c]{@{}l@{}}\textbf{0.289}\end{tabular}} & {\color[HTML]{000000} \begin{tabular}[c]{@{}l@{}}\textbf{0.342}\end{tabular}} & {\color[HTML]{000000} \begin{tabular}[c]{@{}l@{}}\textbf{0.394}\end{tabular}} & {\color[HTML]{000000} \begin{tabular}[c]{@{}l@{}}\textbf{0.179}\end{tabular}} & {\color[HTML]{000000} \begin{tabular}[c]{@{}l@{}}\textbf{0.297}\end{tabular}} & {\color[HTML]{000000} \begin{tabular}[c]{@{}l@{}}\textbf{0.596}\end{tabular}}\\
 \bottomrule
\end{tabular}
\end{table}

\section{Conclusion and Future Work}
\label{sec:conclusions}

We present \emph{WikidataRec}, a hybrid recommender model that recommends Wikidata items to editors based on their past editing activities. The work is motivated by Wikidata's quest for more (and more engaged) editors to keep up with a knowledge graph of growing size and complexity. As the first work of its kind, our focus is on establishing technical feasibility and providing a benchmark for future recommendation research in Wikidata. We do so with solid system implementation and benchmark datasets. Our model is informed by related research in content-based and collaborative filtering in similar verticals, which cannot rely upon explicit feedback for the recommendation task. It uses state-of-the-art models for representation learning and operates by means of a mixture of experts \cite{jacobs1991adaptive}. We employ ELMo \cite{peters2018deep} for items' content-based representations and TransR \cite{lin2015learning} for items' relations-based representations. The results, though far from perfect, are promising and could be considered a baseline for future Wikidata recommender work. 

Based on our experiments, we make three claims for the recommendation task in Wikidata: 1) Collaborative filtering is not enough to recommend Wikidata items, as editing data is very sparse; however, adding item content and relational representations significantly enhances performance; 2) Not each item representation contributes to the final predictions equally. We show how to optimise the weights with NMoR; 3) While our model works better on the denser datasets extracted from \emph{Wikidata-Active-Editors} data, showing that there is room for improvement.


We plan to extend the work in several directions. First, we plan to run editor studies, including interviews with more or less experienced editors to learn about their current ways to choose what they work on and perhaps uncover how existing technical affordances and interfaces influence such decisions. Second, we aim to experiment with other recommender models, particularly sequential learning and time-sensitive models that encode the temporal aspect and distinguish between older and more recent editor interests, as well as with ways to elicit more information about the editors in a responsible way. One option here would be to recruit new editors for a study in which they would explicitly consent to us collecting such information or would provide additional information about their interests themselves. Also, since the data of Wikidata can be represented as item-item relation graph and editor-item graph, exploring graph neural networks (GNNs) for a recommendation would be an interesting future direction. The advantages provided by the GNNs would provide great potential to advance our recommendation task. Third, topic-recommendation is another area that might interests the editors in Wikidata more than item recommendations. We plan to investigate this area. In addition, we are going to conducting an editor-centric evaluation would provide a space to examine and compare the feasibility and effectiveness of the two algorithms (topic vs. item recommendations).



\bibliographystyle{splncs04}
\bibliography{sample}

\end{document}